\documentclass[conference]{IEEEtran}
\IEEEoverridecommandlockouts
\usepackage[english]{babel}
\usepackage{amsmath}
\usepackage{float}

\usepackage[table]{xcolor}
\usepackage{soul}
\usepackage{xfrac}
\usepackage{enumitem}
\usepackage{hyperref}
\hypersetup{final,colorlinks=true}
\usepackage[capitalise]{cleveref}
\usepackage{subcaption}
\usepackage[normalem]{ulem} 

\definecolor{darkgreen}{rgb}{0,0.4,0}
\newcommand*{\numero}{n\kern-.1em \raise.7ex\vbox{\hbox{\tiny \ensuremath{\circ}}\kern.5pt}}

\usepackage{listings}

\setlist[description]{leftmargin=0.1in}
\setlist[itemize]{leftmargin=0.1in}
\begin{document}

\title{Software Defined Networks Key Relay for Large-Scale Quantum Key Distribution Networks}
\author{
\IEEEauthorblockN{Stephan Laschet\IEEEauthorrefmark{1}}
\IEEEauthorblockA{\small ORCID: 0009-0003-1142-8870}

\and
\IEEEauthorblockN{Gergely Lendvay\IEEEauthorrefmark{1}}
\IEEEauthorblockA{\small ORCID: 0009-0003-9228-189X}

\and
\IEEEauthorblockN{Thomas Lorünser\IEEEauthorrefmark{1}}
\IEEEauthorblockA{\small ORCID: 0000-0002-1829-4882}

\and
\IEEEauthorblockN{Paul James\IEEEauthorrefmark{1}}
\IEEEauthorblockA{\small ORCID: 0009-0008-7046-2151}

\and
\IEEEauthorblockN{Luca Torresetti\IEEEauthorrefmark{1}}
\IEEEauthorblockA{\small ORCID: 0009-0002-1257-061X}

\and
\IEEEauthorblockN{Alessandro Colombo\IEEEauthorrefmark{1}}
\IEEEauthorblockA{\small ORCID: 0009-0009-4574-7101}

\and
\IEEEauthorblockA{
\IEEEauthorrefmark{1}Center for Digital Safety and Security,\\
AIT Austrian Institute of Technology, Vienna, Austria
}
}

\maketitle

\begingroup
\renewcommand\thefootnote{}
\footnotetext{\scriptsize
© 2026 IEEE. Personal use of this material is permitted.
Permission from IEEE must be obtained for all other uses, in any
current or future media, including reprinting/republishing this
material for advertising or promotional purposes, creating new
collective works, for resale or redistribution to servers or lists,
or reuse of any copyrighted component of this work in other works.
Published version: \url{https://doi.org/10.1109/QCNC69040.2026.00115}}
\addtocounter{footnote}{-1}
\endgroup

\begin{abstract}
This work addresses the orchestration of large-scale Quantum Key Distribution Networks (QKDNs) using Software Defined Networking (SDN).
Building on ETSI and ITU specifications, common best practices and architectures are outlined. 
The main task of the SDN Controller is to aggregate technical key performance indicators (KPI) from the network and, based on these, select the optimal path.
Multiple path selection algorithms, based on Dijkstra or a maximum-minimum capacity algorithm, with built-in load balancing are presented.
The algorithms were tested in simulations and their performances, and tradeoffs, are discussed.
Additional critical aspects related to SDN controlled QKDNs are discussed, such as query batching, multi-path selection and group key capabilities.
An oblivious multi-party protocol is proposed for relay path selection in a multi-domain scenario, so providers don't have to disclose sensitive information about their QKDN. 
These contributions aim to enhance scalability, resilience and interoperability in quantum-secure network infrastructures.
\end{abstract}

\maketitle

\section{Introduction}\label{sec:intro}

As quantum computer maturity progresses~\cite{gill2025quantum}, it threatens commonly used asymmetric cryptography, since it enables efficient attacks of mathematical cryptographic foundations, using Shor's algorithm~\cite{SHORsAlgo}.
To counteract this threat, one solution is Quantum Key Distribution (QKD)~\cite{bennett2014quantum} which does not only withstand quantum-driven attacks, but also any future technological advancement.
Instead of relying on computational hardness assumptions, QKD establishes a symmetric cryptographic key based on the laws of quantum physics, more specifically, on the non-cloning theorem~\cite{wootters1982single,DIEKS1982271} and Information Theoretically Secure (ITS) algorithms~\cite{ShannonITS}.

But QKD comes with some unique characteristics that need to be accounted for.
The limitation that only directly connected parties within a certain range can establish a key is in stark contrast to the technology QKD is supposed to replace.

To overcome this, QKD links are combined to a QKD Network (QKDN).
The QKD pairs still only establish point-to-point keys, but a Key Management System (KMS) layer above uses them and ITS-algorithms to establish end-to-end keys between any party in the QKDN.
Different techniques exist to relay the keys~\cite{ITU3803}, all requiring a continuous path through the network.
The Software Defined Networking (SDN)~\cite{SDNrfc7426,sdn} approach seems to be one of the most promising candidates for managing QKDNs, many ETSI specifications utilize this technology, as elaborated later.
Not only does SDN enable a holistic network view for relay path selection, it allows additional use-cases.

Especially for multi-domain QKDNs such as EuroQCI~\cite{EuroQCI}, a consistent path calculation in different domains is of upmost importance.
However, as shown in \cref{sec:soa}, those topics are underrepresented in specifications and publications, therefore the authors' findings and approaches are presented in this work.

\section{Contribution}

The state of the art related to architecture, KPIs and path finding is discussed in \cref{sec:soa}.
Given the identified gaps, this work contributes by presenting the authors' findings related to metrics in \cref{sec:kpi} and path finding algorithms in \cref{sec:path_selection}, including results from a simulation framework~\cite{ait-crypto_sdn_simulation}.
This work also addresses the gap in SDN Agent to KMS interface specification by publishing an OpenAPI specification~\cite{ait-crypto_QUICKS}.
Further open issues, such as multi-path and group key support, are discussed in \cref{sec:others}.
Finally, in the light of multi-domain QKDNs, oblivious inter-domain routing is proposed in \cref{sec:OIDR}.

\section{State of the Art and related work}
\label{sec:soa}

\subsection{Related work at ETSI GS QKD}

\begin{figure}[h!]
    \centering
    \includegraphics[width=0.9\columnwidth]{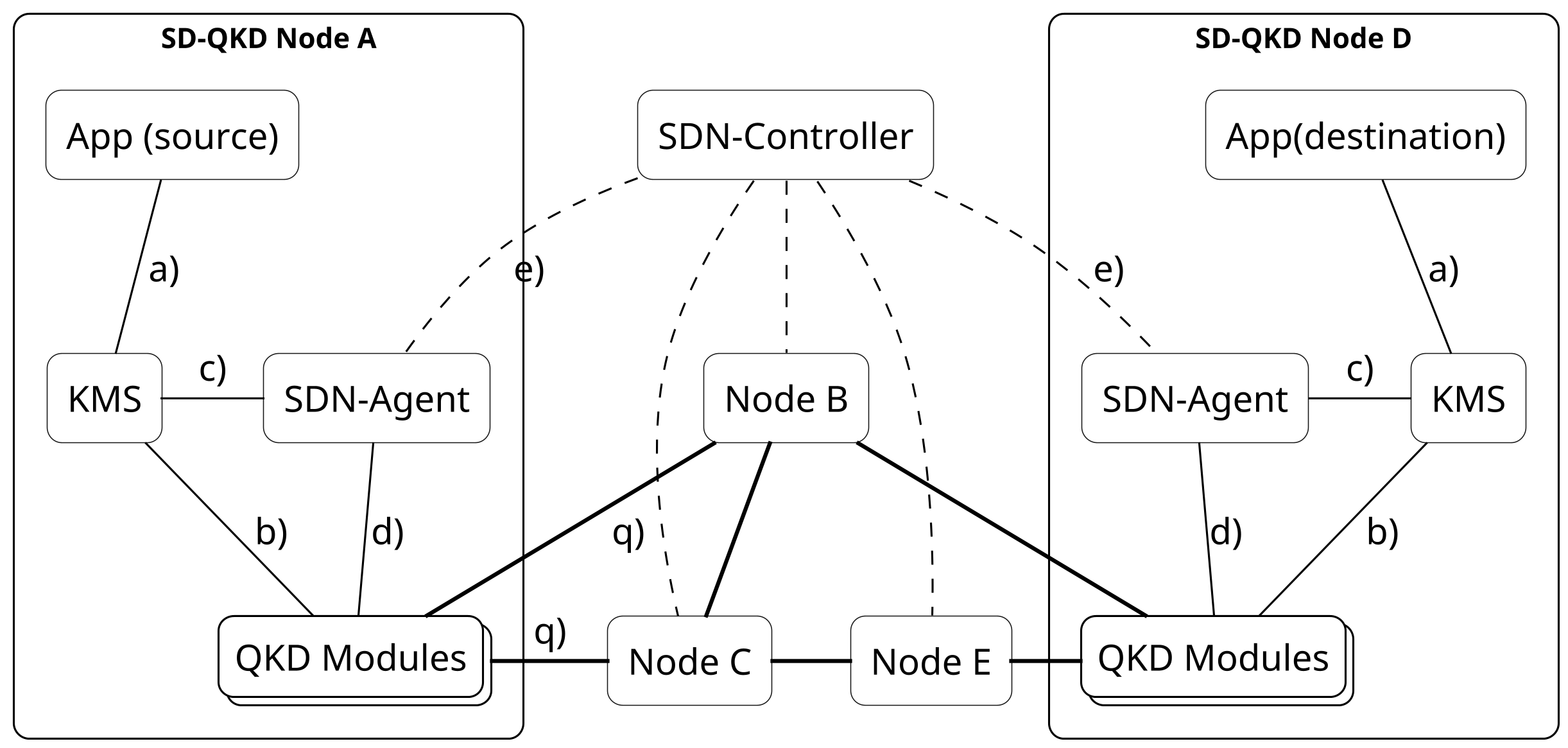}
    \caption{ETSI GS QKD 015~\cite{ETSI015} SDN managed QKDN. Node A and D are fully shown, others compressed.}
    \label{fig:ETSI015_SDN_NW}
\end{figure}

The ETSI GS QKD 015 specification~\cite{ETSI015} focuses on the use of SDN in QKDNs and proposes a high-level architecture as depicted in \cref{fig:ETSI015_SDN_NW}.
It mainly defines an interface between the SDN Controller and the SDN Agent (\cref{fig:ETSI015_SDN_NW}, e).
The SDN Agent represents and models the SD-QKD Node, which is composed of the KMS and QKD modules.
It forwards the SD-QKD Node data to the SDN Controller and receives configuration data from it to distribute within the Node.

Other SDN related ETSI GS QKD specifications are:
\begin{itemize}
  \item ETSI GS QKD 018~\cite{ETSI018} SDN for coexistence of classical and quantum channels.
  \item ETSI GS QKD 021 Draft~\cite{saez2024current} SDN interface to manage separated QKDNs.
  \item ETSI GS QKD 023 Draft~\cite{saez2024current} SDN interface to monitor the KPIs of QKD modules.
\end{itemize}
The SDN Agent to KMS API (\cref{fig:ETSI015_SDN_NW}, c) lacks specification as mentioned in previous publications~\cite{ait_kms}.
To address the lack of specification, the authors' OpenAPI specification~\cite{ait-crypto_QUICKS} is released in the context of this paper.

\subsection{Related work at ITU-T}

The main ITU-T recommendations are Y.3804~\cite{ITU3804} and Y.3805~\cite{ITU3805}, which describe QKDN monitoring and control.

\subsection{QKDN application interfaces}

The application interfaces (\cref{fig:ETSI015_SDN_NW}, a) must be considered for designing the SDN as their behavior significantly determine the behavior of the QKDN.

\subsubsection{ETSI GS QKD 014}

This API~\cite{ETSI014} is commonly used for obtaining keys from the QKDN by an application.
The application states the source and destination as well as the number and size of keys, which is the origin of that data for the SDN.
The API is stateless and does not define a Quality of Service (QoS).
Theoretically an application can ask for any peer as often and for as many keys as it likes.

It also defines a group key feature, for sharing keys with several peers.

\subsubsection{ETSI GS QKD 004}
In contrast, this API~\cite{ETSI004} is stateful using a session with clearly defined QoS, e.g. defining maximum key rates.
Each session is only valid for a dedicated application pair.

\subsubsection{API comparison}

The ETSI GS QKD 014 is more accepted in the industry, since its stateless and unrestricted behavior works well for applications, but makes path selection and service upkeep more challenging.
The ETSI GS QKD 004 is more restrictive to applications, but the session concept works better for SDN path selection and monitoring.

\subsection{Related work on QKDN SDN relay}

Some QKDNs are SDN managed, for example MadQCI~\cite{madqci}, but details about their path finding algorithm are not disclosed, beyond it considering QBER and SKR~\cite{madqci_ba}.
The SeCoQC project~\cite{dianati2008architecture} uses Dijkstra, but doesn't disclose the weighting functions.
The DISCRETION project showed military SDN managed QKDNs, but does not disclose details on path finding algorithms~\cite{discretion}.
The DemoQuanDT project published their path finding approach: They used Dijkstra with a weighting function considering the key store emptiness and the inverse of the SKR~\cite{demoquandt}.
ADA-QKDN uses SPF with the weighting function equivalent to $1/\text{ESKR}$~\cite{ADA_qkdn}.

\section{KPI selection}
\label{sec:kpi}
Technical Key Performance Indicators (KPI) for path selection are discussed in this section.
Mainly the KMS layer is considered, as it performs key relay and interacts with applications.

\subsection{Related work on KPIs for SDN QKDN}
\label{sec:kpi_soa}

The ETSI GS QKD 015~\cite{ETSI015} defines three main KPIs: The \emph{Secure Key Rate (SKR)} is the generation rate of the QKD link. The \emph{Effective Secret Key Rate (ESKR)} is equal to the SKR minus internal key usage. The \emph{Expected consumption} is the key rate queried by applications.

The ITU-T recommendations on QKDN QoS metrics, Y.3806~\cite{ITU3806} and Y.3807~\cite{ITU3807}, focus on QKDN performance for applications, and are less suited for relay path finding.

\subsection{KPI definitions}
\label{sec:kpi_definition}

As the KPIs of \cref{sec:kpi_soa} were found to be suboptimal for path selection, this section introduces other KPIs.
They are vendor independent and easy to determine for any QKDN.

Since each edge in the path must have enough keys stored to support the requested amount, obviously the SDN Controller must be aware of this metric.
The ESKR defined in ETSI GS QKD 015 is impractical as it is a rate, not an absolute value.
Only if the SDN was recording the ESKR for each edge from the beginning, the currently buffered keys could be calculated.

Therefore, the metric \textbf{\emph{Key availability (KAV)}} is introduced.
This metric gives the available key material in bytes on the specified QKD link, which could be delivered to applications.
A KMS can trivially obtain this KPI from its database and publish it to the SDN.
The ESKR can still be determined by calculating $\text{ESKR} = \Delta \text{KAV}/\Delta t$.

In a QKDN, frequently used edges should be avoided, due to the increased load making it more likely that this link runs out of keys.
Hence, the KPI \textbf{\emph{edge usage}} is introduced giving the number of relay paths using a specific edge.
Third, the \textbf{\emph{hop count}} is considered, measuring the number of nodes to be traversed in the specified path.
Overall it is less relevant than in internet routing, however hops add delays, overall more keys are used for the relay process, and each hop exposes the relayed key to potential attacks on trusted nodes.

The most important metric is the KAV, as it is the scarcest resource, determines the feasibility of path establishment and the available reserve for future requests.
The second is the edge usage for load balancing, the third is the path hop count.
To summarize in priority order:

\begin{description}
    \item[Key availability (KAV):] This metric given in bytes determines the amount of key material at the KMS on the direct QKD link available for applications.
    \item[Edge usage:] This metric given as unsigned integer determines, how many paths are using this direct QKD link.
    \item[Hop count:] This metric given as unsigned integer determines how many nodes must be traversed to reach the destination.
\end{description}

Other metrics from the QKDN are still relevant for monitoring and (re-)configuration.

\section{Path selection Algorithms}\label{sec:path_selection}

As discussed in \cref{sec:soa} limited information is published on QKDN path finding, therefore this section outlines the authors' research on the matter.
The authors acknowledge that the effectiveness of a particular algorithm, heuristic, or weighting strategy may vary for different network structures.

Graph theory is used for path selection, where the QKDN is represented by nodes incorporating the QKD, KMS and application instances.
Nodes connected by an edge generate point-to-point QKD keys.
An edge is assigned a weight, so it can be compared to other edges usable for a path.
To generate the weights, the relevant KPIs of \cref{sec:kpi} are compiled by the SDN Controller.
Multiple path selection algorithms and weighting functions were studied, a small selection is discussed in the following sections.

\subsection{Preprocessing and KPI scaling}
\label{sec:preprocessing}

Before calculating the weights, pre-processing is performed.
First, edges with a KAV lower than the amount of requested keys are removed.
Then, it is validated that the source and destination node can still be connected.

To give all KPIs an equivalent contribution they are normed from 1 to 10.
This is done because the KAV is usually in the range of thousands of bytes, while edge usage and hop count are much smaller.
The corresponding maximum value in the QKDN is assigned a 10, the minimum a 1, intermediate values are scaled as rounded integers.
While KAV is always scaled, edge usage and hop count are only scaled if they exceed 10.
\subsection{Dijkstra path finding}
Dijkstra's algorithm~\cite{dijkstra2022note} is a commonly used path selection algorithm and is used in some referenced publications of \cref{sec:soa}.
Dijkstra inherently penalizes high hop counts, as each additional hop increases the total path cost. 
Therefore, we do not consider this KPI in the weighting function.
However, Dijkstra selects the path with minimal weight, while the KAV is defined so that high values are better.
This has to be considered when designing the weighting function.

\subsubsection{Dijkstra weighting function premise}
\label{sec:weighting_func_design}

To guide the formula design, the relation of $\text{KAV}_N$ and $\text{usage}_N$ of $\text{edge}_N$ are compared to the KPIs of $\text{edge}_M$ and finally defined as: 

Edges with a $\text{KAV}_N\leq 3$ should have a higher weight $w$ than edges with the best KAV, regardless of the usage.
An $\text{edge}_N$ with the best $\text{KAV}$ and the highest $\text{usage}$ should be equivalent to an $\text{edge}_M$ with 40\% of the max $\text{KAV}$ and the lowest usage. 
In other words:

\begin{equation}
  \label{equ:premise}
  \begin{gathered}
      \text{$\forall$  } \text{KAV}_N \leq 3: f_N(\text{KAV}_N, \text{usage}_N) \geq f_M(10,\text{usage}_M) \\
      \text{and } f_N(4,1) = f_M(10,10)
  \end{gathered}
\end{equation}

\subsubsection{Linear weighting function}
\label{sec:Dijkstra_lin}

The normed $\text{KAV}_N$ is subtracted from 11, so a higher KAV adds less weight.
A scaling factor increases the KAV impact compared to the usage.
The function is defined as:

\begin{equation}
    \label{equ:weight_lin}
  \begin{gathered}
    w_N = 1.5(11-\text{KAV}_N) + \text{usage}_N
  \end{gathered}
\end{equation}
This algorithm and weighting function combination was dubbed \emph{linear Dijkstra}.
\subsection{Maximum-Minimum Capacity path finding} 
\label{sec:max-min-algo}

The alternative algorithm follows a bottleneck or maximum-minimum capacity approach~\cite{kaymakov2024efficient}, hereafter dubbed \emph{MM-algorithm}.
It selects a path based on the highest minimum weight of an edge in all viable paths between source and destination or in other words the best bottleneck.
The algorithm performs the following steps:
\begin{enumerate}
  \item Obtain all simple paths from source to destination.
  \item Give a score to each edge based on \cref{equ:max_min}.
  \item Based on the worst edge of each path select the path with the highest score.
  \item If there is a tie, repeat selection for the next worst edge.
\end{enumerate}

In contrast to Dijkstra, this approach does not inherently consider the amount of hops in the path, therefore the hop count KPI is included in \cref{equ:max_min}:
\begin{equation}
    w_N = c_1 \cdot \text{KAV}_N - c_2 \cdot \text{usage}_N - c_3 \cdot \text{hops}_N
    \label{equ:max_min}
\end{equation}
where $c_1$, $c_2$ and $c_3$ are constants for scaling.
The constants of $c_1 = 3$, $c_2 = 2$ and $c_3 = 1$ gave the best results and comply with the reasoning given in \cref{sec:kpi_definition}.

\subsection{Comparison and evaluation}

\label{sec:comparison_eval}

While an SDN for QKDN implementation exists at the authors' institute, as described in ~\cite{ait_sdn_bachelor}, the path finding algorithms with different weighting functions were evaluated in a simulation, published on GitHub~\cite{ait-crypto_sdn_simulation}.
This allowed the study of different complex QKDNs and different algorithms.
The simulation assumes the pre-processing of \cref{sec:preprocessing} was already performed.
Evaluation based on the authors' experience from projects mentioned in \cref{sec:ack} was conducted.

While complex graphs representing large scale QKDNs have been studied, to present the observed effects in this short paper, an example graph is discussed.
It has a source (0) and destination (4) and three different paths:

\begin{description}
    \item[direct path]: A direct path, but it has the lowest KAV.
    \item[1 hop]: A path with one intermediate node, but good KAV.
    \item[2 hops]: A path with two intermediate nodes, two edges have the best KAV and one a low KAV.
\end{description}

Considering the priorities of \cref{sec:kpi_definition}, the direct path with the worst KAV should be avoided in this graph.
The path with only one intermediate node and good KAVs seems the best one.
Also, the path with two intermediate nodes is viable due to the best KAV, as hops should matter the least.

\subsubsection{Simulation results}

\begin{figure*}[h]
  \centering

  \begin{subfigure}{0.24\textwidth}
    \centering
    \includegraphics[width=\linewidth]{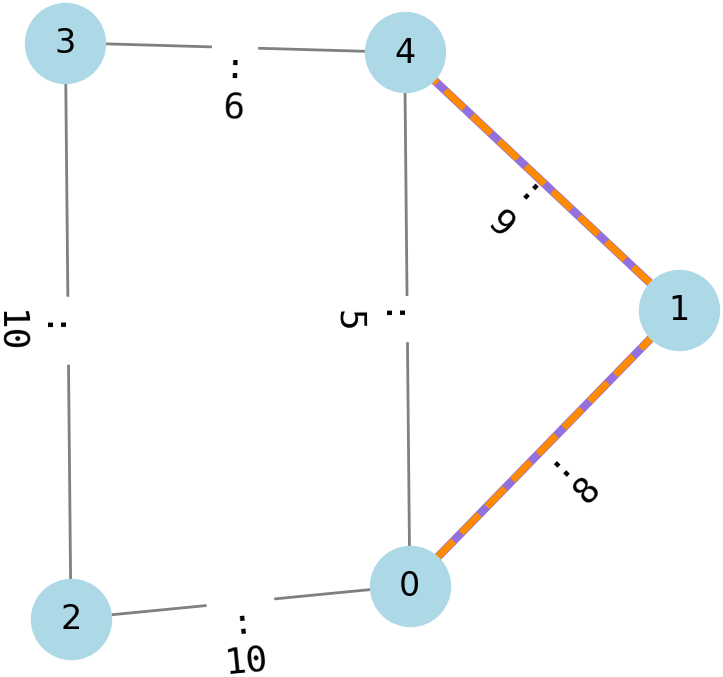}
    \caption{$N = 1$}
  \end{subfigure}
  \hfill
  \begin{subfigure}{0.24\textwidth}
    \centering
    \includegraphics[width=\linewidth]{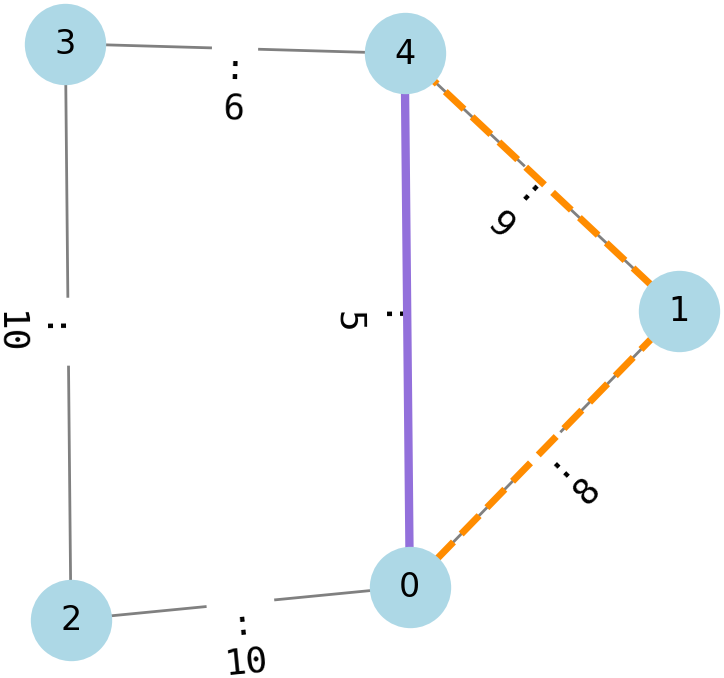}
    \caption{$N = 4$ \& $N = 6$}
  \end{subfigure}
  \hfill
  \begin{subfigure}{0.24\textwidth}
    \centering
    \includegraphics[width=\linewidth]{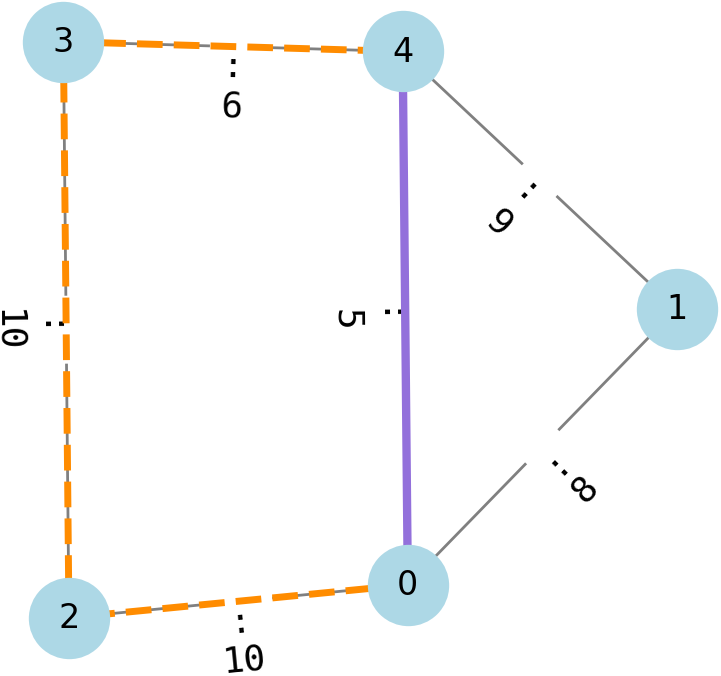}
    \caption{$N = 5$ \& $N = 8$}
  \end{subfigure}
  \hfill
  \begin{subfigure}{0.24\textwidth}
    \centering
    \includegraphics[width=\linewidth]{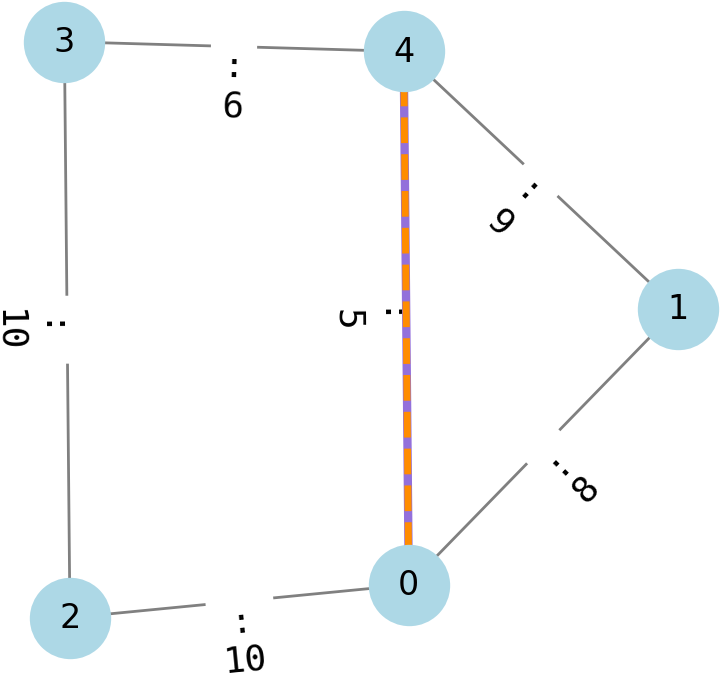}
    \caption{$N = 7$}
  \end{subfigure}
  \caption{Comparison of Dijkstra and MM-algorithm from node 4 to 0, executed $N$ times. The orange dashed path uses the MM-algorithm of \cref{sec:max-min-algo} and the bold purple uses the linear Dijkstra of \cref{sec:Dijkstra_lin}. 
  }
  \label{fig:di_mm_comparison}
\end{figure*}

The simulation results for this example are shown in \cref{fig:di_mm_comparison}, where the MM-algorithm is compared to the linear Dijkstra.
Multiple iterations are shown to determine the influence of increased usage.
In the first 4 iterations the 2 node path is used by both, which is the desired behavior.
Then the linear Dijkstra stays at the direct link.
The MM-algorithm switches to the 3 node path for $N=5$ and 2 node path at $N=6$.
For $N=7$ for the first time the MM-algorithm chooses the direct path.
For higher $N$ alternating the 3 node, 2 node and 1 node path is chosen by the MM-algorithm.

Other weighting functions have been investigated, e.g. ones based on the inverse of the KAV as some publications in \cref{sec:soa} use it.
The detailed analysis exceeds the short paper format, but in short those underperformed as they tend to overemphasize the number of hops.

\subsection{Path selection conclusion}

The linear Dijkstra and MM-algorithm behaved very close to the desired behavior outlined in \cref{sec:comparison_eval}.
The Dijkstra algorithm has the advantage of being efficient and well-studied.
The MM-algorithm was studied as an alternative candidate.
It showed the most reasonable results, but for bigger networks it is more computationally expensive.

\section{Additional considerations for SDN-QKDN}
\label{sec:others}

This section outlines other topics, related to path selection.

\subsection{Key request batching}

To reduce the queries to the SDN Controller for path calculation, a key batching optimization is useful where, instead of establishing one key at a time, a batch of keys is allocated for the application.
Additionally, the internal key usage for the ITS protection should be considered and modeled as a second batch.
So, while the application asks for $N$ keys, the KMS requests allocation of $M$ keys from the SDN with $M = B_{ext} \cdot N + B_{int}$, where $B_{ext}$ is the application batch size multiplier and $B_{int}$ is the internal batch size.
Subsequent application key requests don't require a path selection by the SDN Controller, until the batch is depleted.

\subsection{Multi-Path Selection}
\label{sec:multi-path}

Multi-path key distribution can increase the security in QKDNs~\cite{ait_multipath}, where multiple disjoint paths are chosen.
The SDN Controller has to consider all sets of disjoint paths and determine the optimal ones, based on the KPIs.

ETSI GS QKD 015 does not support this feature and would have to be adapted.
The \texttt{virtual\_link\_spec} must be made a list of \texttt{prev\_hop} and \texttt{next\_hop} instead of single UUID entries.
Additionally, a parameter must define the multi-path algorithm and the number of shares.

\subsection{Group Key Distribution}

The group key is defined in ETSI GS QKD 014 for applications with more than two participants, for example group chats.
The SDN Controller then has to configure multiple paths before the key is distributed in the KMS layer.
Similarly to \cref{sec:multi-path}, for this ETSI GS QKD 015 has to be modified to support multiple \texttt{next\_hop} and \texttt{prev\_hop} values.

\section{Oblivious Inter-Domain Routing}
\label{sec:OIDR}

One of the motivations for this work is to facilitate multi domain QKDNs in the light of EuroQCI.
Therefore, not only a consistent path selection is required, also trustworthiness and collaboration between providers, without disclosing internals, must be ensured.
A QKDN is a very sensitive infrastructure, requiring high security, therefore the network's current status and metrics can't be shared.

Therefore, we propose oblivious inter-domain routing for QKD. 
If domains are connected through some border nodes, we envisage the SDN Controllers negotiating a path obliviously by running secure multi-party protocols~\cite{ITUX1770}.
By leveraging secure computing technologies, a path can be found without revealing any information about the respective network topologies.
This approach maximally preserves the confidentiality of network information while still enabling the SDN to select the globally optimal path.
The linear Dijkstra (\cref{sec:Dijkstra_lin}) is well suited to this problem because it is a greedy algorithm, enabling the providers to pre-compute metrics locally from the source and destination to all border nodes.
However, the concept could also be applied to other routing algorithms.

In our Oblivious Inter-Domain Routing (OIDR) proposal, each SDN controller computes aggregated metrics from each internal node to all available border nodes or between border nodes.
These metrics serve as secure input for the multi-party protocol run between the SDN controllers to evaluate the path selection obliviously.
The oblivious part involves evaluating the sum of the metrics and finding the maximum combination to identify the best path.
If only the optimal solution is revealed, the providers do not learn anything about the other party's network internals, yet the optimal overall path is selected and key availability or link usage does not have to be published.
This approach preserves the confidentiality of network information as much as possible while enabling the selection of the globally optimal path.

\section{Conclusion}\label{sec:conclusion}

This work contributes to SDN for QKDN by identifying gaps in the current specification landscape and proposes technical KPIs and algorithms for path selection.
These algorithms were simulated and the results are compared.
Additional considerations are discussed, such as key batching, multi-path, group keys and oblivious multi-party computation for inter-domain path selection.

\section{Acknowledgment}
\label{sec:ack}
This work received funding from the Österreichische Forschungsförderungsgesellschaft mbH (FFG) program ``Breitband Austria 2030: GigaApp 2. Ausschreibung'', project number (PN): FO999917949 (``Q-Crit Austria''). Also, from the European Union's Horizon Europe research and innovation program PN: 101114043 (``QSNP'') and from the Digital Europe Program, PNs: 101091642 (``QCI-CAT''), 101091588 (``QUARTER''), and 101091564 (``eCausis'').

\bibliographystyle{IEEEtran}
\bibliography{Bibliography}

@incollection{gill2025quantum,
  title     = {Quantum computing: Vision and challenges},
  author    = {Gill, Sukhpal Singh and Cetinkaya, Oktay and Marrone, Stefano and Claudino, Daniel and Haunschild, David and Schlote, Leon and Wu, Huaming and Ottaviani, Carlo and Liu, Xiaoyuan and Machupalli, Sree Pragna and others},
  booktitle = {Quantum Computing},
  pages     = {19--42},
  year      = {2025},
  publisher = {Elsevier}
}

@inproceedings{SHORsAlgo,
  author    = {Shor, P.W.},
  booktitle = {Proceedings 35th Annual Symposium on Foundations of Computer Science},
  title     = {Algorithms for quantum computation: discrete logarithms and factoring},
  year      = {1994},
  volume    = {},
  number    = {},
  pages     = {124-134},
  doi       = {10.1109/SFCS.1994.365700}
}

@article{wootters1982single,
  title     = {A single quantum cannot be cloned},
  author    = {Wootters, William K and Zurek, Wojciech H},
  journal   = {Nature},
  volume    = {299},
  number    = {5886},
  pages     = {802--803},
  year      = {1982},
  publisher = {Nature Publishing Group UK London}
}

@article{DIEKS1982271,
  title   = {Communication by EPR devices},
  journal = {Physics Letters A},
  volume  = {92},
  number  = {6},
  pages   = {271-272},
  year    = {1982},
  issn    = {0375-9601},
  author  = {D. Dieks}
}

@article{ShannonITS,
  author   = {Shannon, C. E.},
  journal  = {The Bell System Technical Journal},
  title    = {Communication theory of secrecy systems},
  year     = {1949},
  volume   = {28},
  number   = {4},
  pages    = {656-715},
  doi      = {10.1002/j.1538-7305.1949.tb00928.x}
}

@article{bennett2014quantum,
  title     = {Quantum cryptography: Public key distribution and coin tossing},
  author    = {Bennett, Charles H and Brassard, Gilles},
  journal   = {Theoretical computer science},
  volume    = {560},
  pages     = {7--11},
  year      = {2014},
  publisher = {Elsevier}
}

@techreport{ETSI004,
  type        = {Group Specification},
  month       = {08},
  year        = {2020},
  title       = {Quantum Key Distribution (QKD); Application Interface},
  number      = {ETSI GS QKD 004 v2.1.1},
  institution = {European Telecommunications Standards Institute (ETSI), Industry Specification Groups(ISG)}
}

@techreport{ETSI014,
  type        = {Group Specification},
  month       = {02},
  year        = {2019},
  title       = {Quantum Key Distribution (QKD); Protocol and data format of REST-based key delivery API},
  number      = {ETSI GS QKD 014 1.1.1},
  institution = {European Telecommunications Standards Institute (ETSI), Industry Specification Groups(ISG)}
}

@techreport{ETSI015,
  type        = {Group Specification},
  month       = {04},
  year        = {2022},
  title       = {Control Interface for Software Defined Networks},
  number      = {ETSI GS QKD 015 v2.1.1},
  institution = {European Telecommunications Standards Institute (ETSI), Industry Specification Groups (ISG)}
}

@techreport{ETSI018,
  type        = {Group Specification},
  month       = {04},
  year        = {2022},
  title       = {Orchestration Interface for Software Defined Networks},
  number      = {ETSI GS QKD 018 v1.1.1},
  institution = {European Telecommunications Standards Institute (ETSI), Industry Specification Groups (ISG)}
}

@inproceedings{saez2024current,
  title        = {Current status, gaps, and future directions in quantum key distribution standards: implications for industry},
  author       = {S{\'a}ez, Juan Morales and Perales, Antonio Pastor and Palancar, Rafael Cant{\'o} and Lopez, Diego R and Chavarria, Jes{\'u}s Folgueira and Ayuso, Vicente Martin and Mendez, Juan Pedro Brito},
  booktitle    = {2024 international conference on Quantum Communications, Networking, and Computing (QCNC)},
  pages        = {341--345},
  year         = {2024},
  organization = {IEEE}
}

@techreport{ITUX1770,
  type        = {Recommendation},
  month       = {10},
  year        = {2021},
  number      = {ITU-T X.1770 v1.0},
  title       = {Technical guidelines for secure multi-party computation},
  institution = {International Telecommunication Union (ITU) Telecommunication Standardization Sector (ITU-T)}
}

@techreport{ITU3803,
  type        = {Recommendation},
  month       = {12},
  year        = {2020},
  number      = {ITU-T Y.3803 v1.0},
  title       = {Quantum key distribution networks - Key management},
  institution = {International Telecommunication Union (ITU) Telecommunication Standardization Sector (ITU-T)}
}

@techreport{ITU3804,
  type        = {Recommendation},
  month       = {4},
  year        = {2025},
  number      = {ITU-T Y.3804 v2.0},
  title       = {Quantum key distribution networks - Control and management},
  institution = {International Telecommunication Union (ITU) Telecommunication Standardization Sector (ITU-T)}
}

@techreport{ITU3805,
  type        = {Recommendation},
  month       = {12},
  year        = {2021},
  number      = {ITU-T Y.3805 v1.0},
  title       = {Quantum key distribution networks - Software-defined networking control},
  institution = {International Telecommunication Union (ITU) Telecommunication Standardization Sector (ITU-T)}
}

@techreport{ITU3806,
  type        = {Recommendation},
  month       = {9},
  year        = {2021},
  number      = {ITU-T Y.3806 v1.0},
  title       = {Quantum key distribution networks - Requirements for quality of service assurance},
  institution = {International Telecommunication Union (ITU) Telecommunication Standardization Sector (ITU-T)}
}

@techreport{ITU3807,
  type        = {Recommendation},
  month       = {2},
  year        = {2022},
  number      = {ITU-T Y.3807 v1.0},
  title       = {Quantum key distribution networks - Quality of service parameters},
  institution = {International Telecommunication Union (ITU) Telecommunication Standardization Sector (ITU-T)}
}

@incollection{dijkstra2022note,
  title     = {A note on two problems in connexion with graphs},
  author    = {Dijkstra, Edsger W},
  booktitle = {Edsger Wybe Dijkstra: his life, work, and legacy},
  pages     = {287--290},
  year      = {2022}
}

@article{kaymakov2024efficient,
  title     = {On efficient algorithms for bottleneck path problems with many sources},
  author    = {Kaymakov, Kirill V and Malyshev, Dmitry S},
  journal   = {Optimization Letters},
  volume    = {18},
  number    = {5},
  pages     = {1273--1283},
  year      = {2024},
  publisher = {Springer}
}

@online{EuroQCI,
  author = {European Commission},
  title  = {The European Quantum Communication Infrastructure (EuroQCI) Initiative},
  month  = {04},
  year   = {2023},
  url    = {digital-strategy.ec.europa.eu/en/policies/european-quantum-communication-infrastructure-euroqci}
}

@article{madqci,
  title     = {MadQCI: a heterogeneous and scalable SDN-QKD network deployed in production facilities},
  author    = {Martin, Vicente and Brito, Juan Pedro and Ort{\'\i}z, Laura and M{\'e}ndez, RB and Buruaga, JS and Vicente, RJ and Sebastian-Lombrana, Alberto and Rincon, David and Perez, Fernando and Sanchez, Cesar and others},
  journal   = {npj Quantum Information},
  volume    = {10},
  number    = {1},
  pages     = {80},
  year      = {2024},
  publisher = {Nature Publishing Group UK London}
}

@article{madqci_ba,
  title       = {Dise{\~n}o e implementaci{\'o}n de m{\'o}dulos de seguridad para la red MadQCI},
  author      = {Fern{\'a}ndez, Axel Abad{\'\i}as},
  advisor     = {Brito M{\'e}ndez, Juan Pedro},
  year        = {2025},
  type        = {Bachelor's thesis},
  institution = {Universidad Polit{\'e}cnica de Madrid},
  school      = {Escuela T{\'e}cnica Superior de Ingenieros Inform{\'a}ticos},
  address     = {Madrid, Spain}
}

@inproceedings{discretion,
  author    = {Bastos, Catarina and Pinto, Francisco and Bacar, Rodrigo and Anjos, Gustavo and Almeida, Margarida and Pinto, Armando Nolasco and Chaves, Ricardo and Dias, Tiago and Afonso, Joana and Calé, Rui and Freitas, Miguel and Maia, Luis and Magalhães, Luis and Muñiz, Alejandro and Cantó, Rafael and Brito, Juan P. and Ballesta, Jesus and Méndez, Ruben B. and Laschet, Stephan and Ramacher, Sebastian and James, Paul and Torresetti, Luca and Piscione, Pietro and Abdulwahed, Ahmed K and Giardina, Pietro and Martin, Vicente and Ortiz, Laura and Pastor, Antonio and Muga, Nelson and Silva, Nuno and López, Diego and Vieira, Margarida and Escribano, Carmen and Mengal, Luis},
  booktitle = {2025 International Conference on Military Communication and Information Systems (ICMCIS)},
  title     = {DISCRETION: First Field Demonstration of a Quantum Enabled SDN in the Context of a Military Exercise},
  year      = {2025},
  volume    = {},
  number    = {},
  pages     = {1-8}
}

@inproceedings{demoquandt,
  title        = {Adapted routing in QKD networks for improved resource utilization},
  author       = {Giemsa, Daniel and Gunkel, Matthias and Johann, Tim and Pachnicke, Stephan and Boehm, Robin and Reuther, Falk},
  booktitle    = {Photonic Networks; 24th ITG-Symposium},
  pages        = {1--4},
  year         = {2023},
  organization = {VDE}
}

@article{ADA_qkdn,
  title     = {ADA-QKDN: A new quantum key distribution network routing scheme based on application demand adaptation},
  author    = {Chen, Li-Quan and Zhao, Meng-Nan and Yu, Kun-Liang and Tu, Tian-Yang and Zhao, Yong-Li and Wang, Ying-Chao},
  journal   = {Quantum Information Processing},
  volume    = {20},
  number    = {9},
  pages     = {309},
  year      = {2021},
  publisher = {Springer}
}

@misc{SDNrfc7426,
  series       = {Request for Comments},
  number       = 7426,
  howpublished = {RFC 7426},
  publisher    = {RFC Editor},
  doi          = {10.17487/RFC7426},
  author       = {Evangelos Haleplidis and Kostas Pentikousis and Spyros Denazis and Jamal Hadi Salim and David Meyer and Odysseas Koufopavlou},
  title        = {{Software-Defined Networking (SDN): Layers and Architecture Terminology}},
  pagetotal    = 35,
  year         = 2015,
  month        = 01
}

@article{sdn,
  author  = {Haji, Saad and Zeebaree, Subhi and Saeed, Rezgar and Ameen, Siddeeq and Shukur, Hanan and Omar, Naaman and Ageed, Zainab and Mahmood, Ibrahim and Yasin, Hajar},
  year    = {2021},
  month   = may,
  pages   = {1--18},
  title   = {Comparison of Software Defined Networking with Traditional Networking},
  volume  = {9},
  journal = {Asian Journal of Computer Science and Information Technology},
  doi     = {10.9734/AJRCOS/2021/v9i230216}
}

@inproceedings{ait_multipath,
  author    = {Valbusa, Federico
               and Lor{\"u}nser, Thomas
               and Spini, Gabriele
               and Laschet, Stephan},
  editor    = {Skopik, Florian
               and Naessens, Vincent
               and De Sutter, Bjorn},
  title     = {Relaxing the Single Point of Failure in Quantum Key Distribution Networks: An Overview of Multi-path Approaches},
  booktitle = {Availability, Reliability and Security},
  year      = {2025},
  publisher = {Springer Nature Switzerland},
  address   = {Cham},
  pages     = {183--200},
  isbn      = {978-3-032-00642-4}
}

@phdthesis{ait_sdn_bachelor,
  title    = {Software Defined Networks for Large-Scale Quantum Key Distribution Networks with Machine Learning Enhancement},
  author   = {Gergely Lendvay},
  year     = {2025},
  month    = may,
  day      = {4},
  language = {English},
  type     = {Bachelor's Thesis},
  school   = {IMC University of Applied Sciences Krems}
}

@article{dianati2008architecture,
  title     = {Architecture and protocols of the future European quantum key distribution network},
  author    = {Dianati, Mehrdad and All{\'e}aume, Romain and Gagnaire, Maurice and Shen, Xuemin},
  journal   = {Security and Communication Networks},
  volume    = {1},
  number    = {1},
  pages     = {57--74},
  year      = {2008},
  publisher = {Wiley Online Library}
}

@inproceedings{ait_kms,
  author    = {James, Paul and Laschet, Stephan and Ramacher, Sebastian and Torresetti, Luca},
  title     = {Key Management Systems for Large-Scale Quantum Key Distribution Networks},
  year      = {2023},
  isbn      = {9798400707728},
  publisher = {Association for Computing Machinery},
  address   = {New York, NY, USA},
  doi       = {10.1145/3600160.3605050},
  booktitle = {Proceedings of the 18th International Conference on Availability, Reliability and Security},
  articleno = {126},
  numpages  = {9},
  location  = {Benevento, Italy},
  series    = {ARES '23}
}

@online{ait-crypto_QUICKS,
  author  = {Laschet, Stephan and James, Paul and Lendvay, Gergely and Torresetti, Luca and Colombo, Alessandro},
  title   = {QUICKS specification, v2.0.0},
  version = {2.0.0},
  year    = {2026},
  url     = {github.com/ait-crypto/kms-sdn-agent-interface-specification},
  note    = {GitHub repository}
}

@online{ait-crypto_sdn_simulation,
  author  = {Laschet, Stephan and Lendvay, Gergely},
  title   = {SDN key relay for QKDN simulation, v1.0.0},
  version = {1.0.0},
  year    = {2026},
  url     = {github.com/ait-crypto/sdn-path-finding-simulation},
  note    = {GitHub repository}
}

\end{document}